\documentclass[prl,reprint,superscriptaddress,preprintnumbers]{revtex4-1}
\pdfoutput=1
\usepackage{amsfonts, amssymb, amsmath}
\usepackage{mathrsfs}
\usepackage{hyperref}
\usepackage{tikz}
\usepackage[table]{}
\usepackage{empheq}

\usepackage{tensor}

\usepackage{color}
\definecolor{dark-gray}{gray}{0.20}
\definecolor{gray}{gray}{0.30}
\definecolor{light-gray}{gray}{0.80}
\definecolor{dark-red}{rgb}{0.7,0,0}
\definecolor{dark-green}{rgb}{0.1,0.4,0}
\definecolor{dark-blue}{rgb}{0.3,0.3,0.7}
\definecolor{light-blue}{rgb}{0.8,0.8,1}
\definecolor{blue}{rgb}{0,0,1}
\definecolor{red}{rgb}{1,0,0}
\definecolor{green}{rgb}{0,1,0}

\usepackage{hyperref}
\hypersetup{
	colorlinks=true,
	linkcolor=dark-blue,
	citecolor=dark-red,
	urlcolor=dark-blue,
	linktoc=page
}

\def\SU{{\rm SU}}

\newcommand{\be}{\begin{equation}}
\newcommand{\ee}{\end{equation}}
\newcommand{\bea}{\begin{eqnarray}}
\newcommand{\eea}{\end{eqnarray}}

\newcommand{\U}{\text{U}}

\begin{document}

\title{The Unreasonable Effectiveness of Higher-Derivative Supergravity\\ in AdS$_4$ Holography}

\date{\today}

\author{Nikolay Bobev}

\affiliation{Instituut voor Theoretische Fysica, KU Leuven, Celestijnenlaan 200D, B-3001 Leuven, Belgium}

\author{Anthony M. Charles}

\affiliation{Instituut voor Theoretische Fysica, KU Leuven, Celestijnenlaan 200D, B-3001 Leuven, Belgium}

\author{Kiril Hristov}

\affiliation{INRNE, Bulgarian Academy of Sciences, Tsarigradsko Chaussee 72, 1784 Sofia, Bulgaria}

\author{Valentin Reys}

\affiliation{Instituut voor Theoretische Fysica, KU Leuven, Celestijnenlaan 200D, B-3001 Leuven, Belgium}

\begin{abstract}
\noindent We study four-derivative corrections to four-dimensional $\mathcal{N}=2$ minimal gauged supergravity and show that they are controlled by two real constants. The solutions of the equations of motion in the two-derivative theory are not modified by the higher-derivative corrections. We use this to derive a general formula for the regularized on-shell action for any asymptotically locally AdS$_4$ solution of the theory and show how the higher-derivative corrections affect black hole thermodynamic quantities in a universal way. We employ our results in the context of holography to derive explicit expressions for the subleading corrections in the large $N$ expansion of supersymmetric partition functions on various compact manifolds for a large class of three-dimensional SCFTs. 
\end{abstract}


\maketitle
%
\section{Introduction}\label{sec:intro}
%
String and M-theory are the natural habitats for the gauge/gravity duality. The gauge theory arises from the low-energy dynamics of D/M-branes which in turn are massive objects that admit an alternative gravitational description.  The gravitational side of the duality is under good calculational control in the classical supergravity limit of string and M-theory which allows for explicit calculations of physical observables in the planar limit of the dual gauge theory. Going beyond this approximation requires calculating higher-derivative corrections to ten- or eleven-dimensional supergravity and delineating their effects on holographic observables. This proves to be technically challenging and our goal here is to bypass some of these difficulties by eschewing the need to work in ten or eleven dimensions and study higher-derivative corrections to four-dimensional supergravity instead. 

We focus on the gravity multiplet of four-dimensional $\mathcal{N}=2$ gauged supergravity which captures the universal dynamics of the energy-momentum multiplet in the dual three-dimensional $\mathcal{N}=2$ SCFT. The propagating degrees of freedom in this theory are the metric, an Abelian gauge field called the graviphoton, and two gravitini. The two-derivative action for the bosonic fields is the Einstein-Maxwell action with a negative cosmological constant. The four-derivative action for this model can be studied using techniques from conformal supergravity. We find that there are only two supersymmetric four-derivative terms in the action which have arbitrary real coefficients, $c_1$ and $c_2$. Despite the non-trivial corrections to the Lagrangian of the theory every  solution of the two-derivative equations of motion (EoM) also solves the four-derivative equations. Moreover the amount of supersymmetry preserved by a given solution is not affected by the four-derivative corrections.

These results prove powerful in the context of holography. In particular they allow for an explicit evaluation of the regularized on-shell action of any solution to the four-derivative supergravity theory which in turn captures the path integral of the dual SCFT. In addition, we show how the presence of the higher-derivative corrections modifies the thermodynamics of black hole solutions in the theory. Our main results for the on-shell action and the black hole entropy can be found in \eqref{eq:i4} and \eqref{eq:s4}, respectively.

While the constants $c_{1,2}$ are free parameters in four-dimensional supergravity they should be uniquely fixed by embedding our model in string or M-theory. In the absence of such an explicit embedding we can appeal to the holographically dual field theory and study its path integral to subleading order in the planar limit. Indeed, this proves to be a fruitful strategy in the context of three-dimensional SCFTs realized on the worldvolume of $N$ M2-branes which are dual to orbifolds of the AdS$_4\times S^7$ background of M-theory. Combining our results for the higher-derivative on-shell action with supersymmetric localization results in the large $N$ limit we find that the partition function of the SCFTs, $Z$, on various compact manifolds has the following explicit form:
\begin{equation}
\label{eq:logZ}
	-\log Z = \pi\,\mathcal{F}\,\bigl[A\,N^{\frac{3}{2}} + B\,N^{\frac{1}{2}}\bigr] - \pi\,(\mathcal{F} - \chi)\,C\,N^{\frac{1}{2}} \,.
\end{equation}
The quantities $(\mathcal{F},\chi)$ depend on the compact three-manifold and are given in Table~\ref{tab:solutions} for various cases of interest, while the constants $(A,B,C)$ depend on the SCFT. For the $\U(N)_{k}\times \U(N)_{-k}$ ABJM theory and the $\mathcal{N}=4$ $\U(N)$ SYM theory coupled to one adjoint and $N_f$ fundamental hypermultiplets, the constants $(A,B,C)$ can be computed explicitly and are given in  Table~\ref{tab:coeff}.
\begin{table}[ht]
	\begin{center}
		\setlength{\tabcolsep}{7pt}
		\begin{tabular}{ c || c | c | c |} \hline
	\multicolumn{1}{|c||}{Theory}	 & $A$ & $B$ & $C$ \\ \hline
		\multicolumn{1}{|c||}{ABJM at level $k$} & $\frac{\sqrt{2 k}}{3}$ & $ -\frac{k^2 + 8}{24 \sqrt{2 k}}$ & $-\frac{1}{\sqrt{2 k}}$  \\ \hline
		\multicolumn{1}{|c||}{$\mathcal{N}=4$ SYM w. $N_f$ fund.} & $\frac{\sqrt{2 N_f}}{3}$ & $ \frac{N_f^2 - 4}{8 \sqrt{2 N_f}}$ & $-\frac{N_f^2 + 5}{6 \sqrt{2 N_f}}$ \\ \hline
		\end{tabular}
	\end{center}
	\caption{The constants in \eqref{eq:logZ} for two classes of SCFTs.}
	\label{tab:coeff}
\end{table}
%
%
\section{Minimal gauged supergravity}
%
In the conformal supergravity formalism, the two-derivative action of four-dimensional minimal gauged~$\mathcal{N}=2$ supergravity is specified by the Weyl multiplet, an auxiliary vector multiplet and an auxiliary hypermultiplet, see \cite{Lauria:2020rhc} for a review. In Euclidean signature, the vector multiplet is related to the reducible combination of a real chiral multiplet~$\mathcal{X}_+$ and a real anti-chiral multiplet~$\mathcal{X}_-$, see~\cite{deWit:2017cle}. The combinations~$(\mathcal{X}_\pm)^2$ can be used to construct a supersymmetric Lagrangian density by means of (anti-)chiral superspace integrals~\cite{Butter:2011sr},
\begin{equation}
\label{eq:superspace-vec}
\mathcal{L}_\mathrm{V} = \tfrac{1}{2}\int d^4\theta\,\mathcal{E}_+\,(\mathcal{X}_+)^2 + \tfrac{1}{2}\int d^4\bar{\theta}\,\mathcal{E}_-\,(\mathcal{X}_-)^2 \, ,
\end{equation}
with~$\mathcal{E}_\pm$ the (anti-)chiral superspace measure. The full two-derivative action of minimal gauged supergravity is obtained by supplementing~\eqref{eq:superspace-vec} with the Lagrangian density for the hypermultiplet, whose explicit form can be found in~\cite{deWit:2017cle}. This multiplet is allowed to transform locally under a~$\U(1)$ subgroup of the $\SU(2)$ R-symmetry, implementing the gauging in the supergravity theory. 

We consider two supersymmetric Lagrangian densities containing higher-derivative couplings. The first is built out of the Weyl multiplet, related to a chiral and an anti-chiral tensor multiplet~$\mathcal{W}^{ab}_\pm$, which can be squared to construct the superspace integrals~\cite{Bergshoeff:1980is}
\begin{equation}
\label{eq:superspace-W2}
\mathcal{L}_{\mathrm{W}^2} = -\tfrac1{64}\int d^4\theta\,\mathcal{E}_+\,(\mathcal{W}^{ab}_+)^2 - \tfrac1{64}\int d^4\bar{\theta}\,\mathcal{E}_-\,(\mathcal{W}^{ab}_-)^2 \, .
\end{equation}
The second is built from the so-called $\mathbb{T}$-log multiplet and contains the supersymmetrization of the Gauss-Bonnet term~\cite{Butter:2013lta}. In superspace notation, it reads
\begin{equation}
\label{eq:superspace-T-log}
\mathcal{L}_{\mathbb{T}\text{log}} = -\tfrac{1}{2}\int d^4\theta\,\mathcal{E}_+\,\Phi_+'\overline{\nabla}^{\,4}\ln\Phi_- + \text{anti-chiral} \, ,
\end{equation}
where~$\Phi_+'$ is a chiral multiplet and~$\Phi_-$ is an anti-chiral multiplet. As shown in~\cite{Butter:2013lta}, when the former is a constant multiplet,~$\Phi_-$ can be identified with~$\mathcal{X}_-$ without loss of generality. On the other hand, in minimal gauged supergravity, identifying~$\Phi_+'$ with a composite chiral multiplet (necessarily of zero Weyl weight~\cite{Butter:2013lta}) leads to terms in the Lagrangian~\eqref{eq:superspace-T-log} with at least six derivatives. We do not consider such terms and therefore can fix~$\Phi'_+ = 1$.

The Lagrangians in \eqref{eq:superspace-vec}, \eqref{eq:superspace-W2}, and \eqref{eq:superspace-T-log} are individually superconformally invariant. The two-derivative Lagrangian can be taken with unit coefficient by simple field redefinitions, and we are left with two arbitrary real coefficients~$c_1$ and~$c_2$ of the four-derivative Lagrangians~\eqref{eq:superspace-W2} and~\eqref{eq:superspace-T-log}, respectively. Furthermore, the bosonic terms in~\eqref{eq:superspace-W2} and~\eqref{eq:superspace-T-log} are related by $\mathcal{L}_{\mathrm{W}^2} + \mathcal{L}_{\mathbb{T}\text{log}} = \mathcal{L}_{\text{GB}}$, where $\mathcal{L}_\text{GB}$ is the Gauss-Bonnet invariant~\cite{Butter:2013lta}. We therefore  can eliminate the~$\mathbb{T}$-log Lagrangian in favor of the more familiar Weyl-squared and Gauss-Bonnet terms and write the superconformal higher-derivative Lagrangian as
\begin{equation}
\label{eq:SCHD}
\mathcal{L}_{\text{HD}} = \mathcal{L}_{2\partial} + (c_1 - c_2)\,\mathcal{L}_{\mathrm{W}^2} + c_2\,\mathcal{L}_{\text{GB}} \, .
\end{equation}

Starting from~\eqref{eq:SCHD}, we obtain the corresponding Lagrangian density in the Poincar\'{e} frame by gauge-fixing the extra symmetries and eliminating the fields that have been introduced to guarantee off-shell closure of the superconformal algebra. The result is an action involving only the dynamical fields of minimal gauged supergravity.
This procedure will be presented in detail in~\cite{bigpaper}. Here we make the following key observation: upon choosing convenient gauge-fixing conditions for the superconformal symmetries, the extra superconformal fields can be eliminated from~\eqref{eq:SCHD} using their \emph{two-derivative} solutions, even in the presence of the higher-derivative couplings. The result from this foray into conformal supergravity is the following form of the bosonic terms of~\eqref{eq:SCHD} in the Poincar\'{e} frame:
\begin{align}
\label{eq:PHD}
e^{-1}\mathcal{L}_{2\partial} =&\; -(16\pi\,G_N)^{-1}\bigl[R + 6\,L^{-2} - \tfrac14\,F_{ab}F^{ab}\bigr]\, , \nonumber \\
e^{-1}\mathcal{L}_{\mathrm{W}^2} =&\; \bigl(C_{ab}{}^{cd}\bigr)^2 - L^{-2} F_{ab}F^{ab} + \tfrac12\bigl(F_{ab}^+\bigr)^2\bigl(F_{cd}^-\bigr)^2 \nonumber \\
&- 4\,F_{ab}^- R^{ac} F^+_c{}^b + 8\,\bigl(\nabla^a F_{ab}^-\bigr)\bigl(\nabla^c F^+_c{}^b\bigr) \, , \\
e^{-1}\mathcal{L}_\text{GB} =&\; R^{abcd}\,R_{abcd} - 4\,R^{ab} R_{ab} + R^2 \, , \nonumber
\end{align}
where $G_N$ is the Newton constant, $C_{ab}{}^{cd}$ is the Weyl tensor,~$F_{ab}$ is the graviphoton field strength, and~$L$ determines the cosmological constant. Using~\eqref{eq:PHD}, one can show that the two-derivative solutions for the metric and graviphoton also solve the four-derivative equations of motion, but not vice-versa. Similar results have been noticed also in the context of ungauged~$\mathcal{N}=2$ supergravity~\cite{Charles:2016wjs, Charles:2017dbr}, as well as for non-supersymmetric theories \cite{Smolic:2013gz}. We also find that the Poincar\'{e} supersymmetry variations of the gravitini are not modified by the presence of the higher-derivative couplings. We thus conclude that any solution of the two-derivative Poincar\'{e} action is also a solution of the higher-derivative action~\eqref{eq:SCHD}, and it preserves the same amount of supersymmetry.
%
\section{On-shell action}
In order to evaluate the action \eqref{eq:SCHD} on solutions of the two-derivative EoM we use \eqref{eq:PHD} to derive the following identity for the on-shell values of the three actions:
\begin{equation}
\label{eq:W2-os}
I_{\mathrm{W}^2} = I_\text{GB} -\frac{64\pi G_N}{L^2}\,I_{2\partial} \, .
\end{equation}
The divergences in the on-shell actions on the right-hand side can be removed via holographic renormalization using the following counterterms~\cite{Emparan:1999pm,Myers:1987yn}:
\begin{align}
\label{eq:CTs}
I^\text{CT}_{2\partial} =&\; (8\pi\,G_N)^{-1}\int d^3 x \sqrt{h}\,\bigl( - K + \tfrac{1}{2}\,L\,\mathcal{R} + 2\,L^{-1}\bigr) \, , \nonumber \\
I^\text{CT}_{\text{GB}} =&\; 4\int d^3 x \sqrt{h}\,\bigl(\mathcal{J} - 2\,\mathcal{G}_{ab}\,K^{ab}\bigr) \, ,
\end{align}
where~$h_{ab}$ is the induced metric on the boundary,~$K_{ab}$ is the extrinsic curvature,~$\mathcal{R}$ and $\mathcal{G}_{ab}$ are the boundary Ricci scalar and Einstein tensor, respectively, and~$\mathcal{J}$ is defined by
\begin{equation}
\mathcal{J} = \tfrac13\bigl(3 K (K_{ab})^2 - 2 (K_{ab})^3 - K^3 \bigr) \, .
\end{equation}
Using  \eqref{eq:PHD} and \eqref{eq:CTs}, we can compute the regularized on-shell actions
\begin{equation}
\begin{split}
 I_{2\partial} + I^\text{CT}_{2\partial} = \frac{\pi L^2}{2 G_N}\mathcal{F}  \, , \qquad I_\text{GB} + I^\text{CT}_{\text{GB}} = 32\pi^2\chi  \, ,
\end{split}
\label{eq:regs}
\end{equation}
where~$\mathcal{F}$ depends on the two-derivative solution~$\mathcal{M}_4$, and~$\chi$ is the Euler characteristic of~$\mathcal{M}_4$, see Table~\ref{tab:solutions}. Combining \eqref{eq:SCHD}, \eqref{eq:W2-os}, and \eqref{eq:regs}, we arrive at the following universal formula for the regularized four-derivative on-shell action in minimal gauged supergravity:
\begin{empheq}[]{equation}
\label{eq:i4}
I_{\text{HD}} = \Bigl[1 + \frac{64\pi G_N}{L^2}(c_2 - c_1)\Bigr]\frac{\pi L^2}{2 G_N}\mathcal{F} + 32\pi^2 c_1\,\chi\, .
\end{empheq}
This remarkably simple formula relates the full four-derivative on-shell action to the two-derivative result, proportional to ${\cal F}$, along with the topological invariant~$\chi$. We emphasize that this result for the on-shell action is valid for all solutions of the two-derivative EoM and is independent of supersymmetry. In the context of holography the on-shell action in \eqref{eq:i4} should be dual to the logarithm of the partition function of a three-dimensional $\mathcal{N}=2$ SCFT defined on the boundary of $\mathcal{M}_4$. Standard examples for~$\mathcal{M}_4$ include Euclidean AdS$_4$ solutions with squashed $S^3$ boundary as well as  Euclidean black hole solutions with $S^1 \times \Sigma_{\mathfrak{g}}$ boundary. We present the results for~$\cal{F}$ and~$\chi$ for a number of supersymmetric solutions of minimal gauged supergravity in Table~\ref{tab:solutions}. 
\begin{table}[ht]
	\begin{center}
	\setlength{\tabcolsep}{7pt}
	\begin{tabular}{ c || c| c | c | c |} \hline
	\multicolumn{1}{|c||}{Solution $\mathcal{M}_4$} & Susy	 & Ref. & ${\cal F}$ & $\chi$ \\ \hline \hline
	\multicolumn{1}{|c||}{AdS$_4$ w. $\!S^3$ bdry }& 1 & \cite{Chamblin:1998pz} & $1$ & $1$  \\ \hline
	\multicolumn{1}{|c||}{$\U(1) \times \U(1)$ sq.} & 1/2 & \cite{Martelli:2011fu} & $ \frac14 ( b + \frac{1}{b})^2 $ & $1$ \\ \hline
	\multicolumn{1}{|c||}{$\SU(2) \times \U(1)$ sq.} & 1/2 & \cite{Martelli:2011fw} & $s^2$ & $1$ \\ \hline
	\multicolumn{1}{|c||}{$\SU(2) \times \U(1)$ sq.}& 1/4  & \cite{Farquet:2014kma} & $1$ & $1$ \\ \hline
	\multicolumn{1}{|c||}{KN-AdS} & 1/4 & \cite{Caldarelli:1998hg} & $ \frac{(\omega + 1)^2}{2\omega} $ & $2$ \\ \hline\hline
	\multicolumn{1}{|c||}{AdS$_2 \times \Sigma_{\mathfrak{g}}$} & 1/2 & \cite{Caldarelli:1998hg} & $(1-\mathfrak{g})$ & $2(1-\mathfrak{g})$ \\ \hline
	\multicolumn{1}{|c||}{Romans} & 1/4 & \cite{Bobev:2020pjk} & $(1-\mathfrak{g})$ & $2(1-\mathfrak{g})$ \\ \hline
	\multicolumn{1}{|c||}{Bolt$_{\pm}$} & 1/4 & \cite{Toldo:2017qsh} & $(1-\mathfrak{g}) \mp \frac{\mathfrak{}p}{4}$ & $2(1-\mathfrak{g})$ \\ \hline
	\end{tabular}
	\end{center}
	\caption{The on-shell action~\eqref{eq:i4} for various supersymmetric Euclidean solutions of holographic interest along with references with the explicit form of the metric and gauge field. The double line separates solutions with NUTs and Bolts. }
	\label{tab:solutions}
\end{table}

There are two types of contributions to the on-shell action of supersymmetric solutions, which can be classified by the dimension of the fixed loci under the isometry required by supersymmetry~\cite{BenettiGenolini:2019jdz}.  As shown in~\cite{BenettiGenolini:2019jdz}, the two-derivative on-shell action $\mathcal{F}$ localizes on the fixed points of the preserved equivariant supercharges, and this principle allows for its explicit evaluation for generic NUT or Bolt solutions.  Our result \eqref{eq:i4} demonstrates that the higher-derivative corrections to the on-shell action can also be written purely in terms of topological fixed point data.  It should therefore be possible to repeat the analysis of \cite{BenettiGenolini:2019jdz} for the most general supersymmetric solutions of the four-derivative action in \eqref{eq:SCHD} and establish a higher-derivative generalization of their localization results.

%
%
%
%

Our results can be related to another important observable in AdS$_4$ holography: the coefficient, $C_T$, of the two-point function of the energy momentum tensor in the dual SCFT. Using the four-derivative action in \eqref{eq:PHD} and the results in \cite{Sen:2014nfa} we find (in the conventions of \cite{Chester:2020jay})
\begin{equation}\label{eq:CTSS}
C_T = \frac{32L^2}{\pi G_N} + 2048 (c_2-c_1)\,.
\end{equation}
This result is valid for all three-dimensional holographic SCFTs captured by our minimal supergravity setup. One can use \eqref{eq:i4} and the second entry in Table~\ref{tab:solutions} to confirm the Ward identity~$C_T = \frac{32}{\pi^2} \frac{\partial^2 I_{S^3_b}}{\partial b^2}|_{b=1}$,~see~\cite{Closset:2012ru}. This constitutes a non-trivial consistency check of our results.
%
\section{Black hole thermodynamics}
%
To understand how the thermodynamics of black holes is modified by the four-derivative terms in \eqref{eq:PHD} we consider a stationary black hole solution to the two-derivative equations of motion. To this end we work in Lorentzian signature as implemented via a Wick-rotation of~\eqref{eq:PHD}.

In a higher-derivative theory of gravity, the black hole entropy can be computed using the Wald formalism \cite{Wald:1993nt}:
\begin{equation}
	S = - 2\pi \int_H E^{abcd}\,\varepsilon_{ab}\,\varepsilon_{cd}\, ,
\end{equation}
where the integral is over the two-dimensional horizon~$H$,~$E^{abcd}$ is the variation of the Lagrangian with respect to the Riemann tensor, and~$\varepsilon_{ab}$ is the unit binormal to the horizon. Using the Lagrangian~\eqref{eq:SCHD} and the EoM, we obtain
\begin{equation}
\label{eq:s4}
S = \left(1 + \alpha \right)\frac{A_H}{4G_N} - 32\pi^2 c_1\,\chi(H) \, ,
\end{equation}
where ~$A_H$ is the area and~$\chi(H)$ the Euler characteristic of the horizon and we defined $\alpha:=\frac{64\pi G_N}{L^2}(c_2 - c_1)$. We find two modifications to the entropy: a topological term independent of the charges of the black hole, accompanied by an overall rescaling of the Bekenstein-Hawking area law.  This rescaling can be interpreted as a renormalization of $G_N$, as required to tame divergences in a UV-complete theory of gravity~\cite{Susskind:1994sm,Larsen:1995ax}.

The four-derivative terms in the action modify the conserved quantities associated with Killing vectors of the spacetime.  Let $\Sigma$ be a time-like boundary at spatial infinity. The conserved charge $\mathcal{Q}$ associated with a Killing vector $K$ can be computed by the Komar integral
\begin{equation}
	\mathcal{Q}[K] = \int_{\partial \Sigma} d^2x\,\sqrt{\gamma}\, n^a K^b \tau_{ab}~, \quad \tau_{ab} := \frac{2}{\sqrt{h}} \frac{\delta \mathcal{L}_\text{HD}}{\delta h^{ab}}~,
\label{eq:QK}
\end{equation}
with $\gamma$ the induced metric on the boundary surface $\partial \Sigma$, $n^a$ the unit normal to $\partial \Sigma$, and $\tau_{ab}$ the boundary stress tensor~\cite{Balasubramanian:1999re}. 
Using~\eqref{eq:W2-os} and~\eqref{eq:CTs}, we find that for any solution of the two-derivative EoM the boundary stress tensor takes the universal form
\begin{equation}
	\tau^{ab} = (1 + \alpha) \tau^{ab}_{2\partial} - c_1\,\tau^{ab}_\text{GB}\,,
\end{equation}
where $\tau^{ab}_{2\partial}$ is the boundary stress-tensor associated with $\mathcal{L}_{2\partial}$ and $\tau^{ab}_\text{GB}$ is the boundary stress tensor associated with $\mathcal{L}_\text{GB}$ in \eqref{eq:PHD}.  Crucially, the topological nature of the Gauss-Bonnet term ensures that $\tau^{ab}_\text{GB}$ gives no contribution to the Komar integral~\cite{Myers:1988ze}.  This implies that the four-derivative terms in \eqref{eq:PHD} simply rescale the Komar charges of the original two-derivative solution.  In particular, the mass and angular momentum of the black hole are $M = (1 + \alpha) M_{2\partial}$ and $J = (1 + \alpha) J_{2\partial}$, respectively. 

To study the electromagnetic charges of the black hole we note that the Maxwell equations can be written as $d G = d F = 0$, where $F$ is the two-form graviphoton field strength and $G$ is the two-form defined by
\begin{equation}
	(\star G)_{\mu\nu} = 32 \pi\,G_N\,\frac{\delta \mathcal{L}_\text{HD}}{\delta F^{\mu\nu}}\,.
\end{equation}
The electric and magnetic charges $Q$ and $P$ are defined by integrating $G$ and $F$ over $\partial\Sigma$:
\begin{equation}
	Q = \int_{\partial\Sigma} G~, \qquad P = \int_{\partial\Sigma} F\,.
\end{equation}
The field strength~$F$, and therefore the magnetic charge, is unaffected by the higher-derivative terms. However, the latter modify $G$, which in turn modifies the electric charge as $Q = (1 + \alpha) Q_{2\partial}$. 

As a consistency check of our results we consider the quantum statistical relation between the thermodynamic quantities of a black hole and its Euclidean on-shell action \cite{Gibbons:2004ai}
\begin{equation}
	I = \beta \left( M - TS - \Phi Q - \omega J\right)~,
\end{equation}
where $T = \beta^{-1}$ is the temperature, $\Phi$ is the electric potential, and $\omega$ is the angular velocity of the black hole. These intensive quantities are fully determined by the two-derivative solution and are therefore not modified since the black hole background is not affected by the four-derivative terms in the action.  The same is not true for the extensive quantities $I$, $S$, $M$, $Q$, and $J$ computed above. Comparing~\eqref{eq:i4} to~\eqref{eq:s4}, we find that if the quantum statistical relation is satisfied in the two-derivative theory then it is automatically satisfied in the four-derivative theory provided that the Euler characteristics of the full Euclidean solution and the horizon are equal, $\chi (\mathcal{M}_4) = \chi(H)$.  We have checked this relation for all known asymptotically AdS$_4$ stationary black holes and it would be interesting to prove it in full generality. 

Our results imply that the ratio $Q/M$ for extremal black holes is not affected by the four-derivative terms in  \eqref{eq:PHD} and thus the corrections to the black hole entropy in \eqref{eq:s4} have no relation to the extremality bound. Moreover, the black hole entropy corrections do not have a definite sign and therefore do not necessarily lead to an increase in the entropy for all black holes. These results are in conflict with some of the claims in \cite{Cheung:2018cwt,Goon:2019faz,Cremonini:2019wdk} and will be discussed further in \cite{bigpaper}.

%
\section{Field theory and holography}
%
The discussion so far was confined to four-dimensional supergravity. To make a connection with holography we now assume that the four-dimensional supergravity action in \eqref{eq:SCHD} arises as a consistent truncation of M-theory on an orbifold of $S^7$. This consistent truncation has been established at the two-derivative level and in the absence of orbifolds \cite{Gauntlett:2007ma}. It will be most interesting to extend this result to include higher-derivative terms and to study potential subtleties arising from orbifolds with fixed points. We consider two classes of orbifolds for which the low energy dynamics of $N$ M2-branes is captured by the $\U(N)_{k}\times \U(N)_{-k}$ ABJM theory \cite{Aharony:2008ug} or a $\U(N)$ $\mathcal{N}=4$ SYM theory with one adjoint and $N_f$ fundamental hypermultiplets \cite{Mezei:2013gqa}.

For theories arising from M2-branes it is expected that the dimensionless ratio $\frac{L^2}{2G_N} $ scales as $N^{\frac{3}{2}}$ while the four-derivative coefficients $c_{1,2}$ scale as $N^{\frac{1}{2}}$, see for example \cite{Camanho:2014apa}. We expect that the coefficients of the six- and higher-derivative terms in the four-dimensional supergravity Lagrangian are  more subleading in the large $N$ limit. To this end it is convenient to define the constants $v_i := 32\,\pi\,c_i\,N^{-\frac{1}{2}}$. In addition, we allow for an $N^{\frac{1}{2}}$ correction to $\frac{L^2}{2G_N} $ by defining
\begin{equation}
\label{eq:q-holo-dict}
	\frac{L^2}{2G_N} = A\,N^{\frac{3}{2}} + a\,N^{\frac{1}{2}} \, .
\end{equation}
With this at hand the on-shell action in~\eqref{eq:i4} becomes
\begin{equation}
\label{eq:iABC}
	I_{\text{HD}} = \pi\,\mathcal{F}\,\left[A\,N^{\frac{3}{2}} + B\,N^{\frac{1}{2}}\right] - \pi\,(\mathcal{F} - \chi)\,C\,N^{\frac{1}{2}} \, ,
\end{equation}
where~$B := a + v_2$ and~$C := v_1$. To determine the constants $(A,B,C)$ we can use supersymmetric localization results on the squashed sphere $S^3_{b}$ corresponding to the second entry in Table~\ref{tab:solutions}. In particular for the round sphere at $b=1$ the free energy for the ABJM theory and the $\U(N)$ $\mathcal{N}=4$ SYM was computed in \cite{Marino:2011eh,Fuji:2011km} and ~\cite{Mezei:2013gqa}, respectively. These results allow us to determine the constant $A$ as well as the sum $B+C$ in \eqref{eq:iABC}. For both families of SCFTs it is also possible to compute the constant $C_T$ \cite{Chester:2020jay} and one can combine this with \eqref{eq:iABC} and the supergravity result in \eqref{eq:CTSS} to determine $B$ and $C$ individually. The outcome of these calculations is summarized in Table~\ref{tab:coeff}. Note that these results unambiguously fix the coefficient $c_1$ in \eqref{eq:SCHD}, while $c_2$ cannot be fully determined due to the shift by the constant $a$ in \eqref{eq:q-holo-dict}.

Using this amalgam of four-derivative supergravity and supersymmetric localization results we arrive at the general form of the partition function for these two classes of SCFTs \eqref{eq:logZ}. As a further consistency check we note that our results for the ABJM theory at level $k=1$ and the $\U(N)$ $\mathcal{N}=4$ SYM theory for general $N_f$ agree with \cite{Hatsuda:2016uqa} where the squashed sphere partition function was computed for $b^2=3$. For more general values of the squashing parameter we obtain the following result for the ABJM free energy, $F := -\log Z$:
\begin{equation}\label{eq:FS3bABJM}
\begin{split}
F_{S^3_b} = \tfrac{\pi\sqrt{2k}}{12}\left[\left(b+\tfrac{1}{b}\right)^2\left(N^{\frac{3}{2}}+\left(\tfrac{1}{k}-\tfrac{k}{16}\right)N^{\frac{1}{2}}\right)   - \tfrac{6}{k}N^{\frac{1}{2}}\right]\,. \notag
\end{split}
\end{equation}
Additionally, the result in \eqref{eq:logZ}, together with the second to last entry in Table~\ref{tab:solutions}, allows for the calculation of  the leading correction to the large $N$ results for the topologically twisted index on $S^1\times \Sigma_{\mathfrak{g}}$ for the so-called universal twist \cite{Benini:2015eyy,Azzurli:2017kxo}. For the ABJM theory we find
\begin{equation}\label{eq:Zs1sigABJM}
\begin{split}
-\log Z_{S^1\times \Sigma_{\mathfrak{g}}} = (1-\mathfrak{g})\tfrac{\pi\sqrt{2k}}{3}\left(N^{\frac{3}{2}} -\tfrac{32+k^2}{16k} N^{\frac{1}{2}}\right)\,.\notag
\end{split}
\end{equation}
This agrees with the result from supersymmetric localization for $\mathfrak{g}=0$ in \cite{Liu:2017vll}. Finally, we note that using the explicit results for $(A,B,C)$ in Table~\ref{tab:coeff} and the result in~\eqref{eq:s4} we can compute the leading correction to the entropy of any asymptotically AdS$_4 \times S^7$ black hole.
%
\section{Discussion}\label{sec:conclusion}
%
We studied four-derivative corrections to minimal $\mathcal{N}=2$ gauged supergravity and analyzed their effects on black hole thermodynamics and holography. It is important to generalize our construction to matter-coupled $\mathcal{N}=2$ gauged supergravity as well as to theories with $\mathcal{N}=4$ and $\mathcal{N}=8$ supersymmetry. In addition, it should be possible to extend our results to minimal gauged supergravity in higher dimensions. Some of these generalizations will be explored in \cite{bigpaper}.

It would be most interesting to derive the supergravity action in \eqref{eq:SCHD} from a KK reduction of type II or eleven-dimensional supergravity with higher derivative corrections. This will allow for a first principle derivation of the constants $c_{1,2}$. In the context of M-theory the relation between supersymmetric localization and higher-derivative holography we explored should be viewed as complementary to the strategy pursued in \cite{Chester:2018aca}. It will be interesting to understand whether we can use this alternative vantage point to determine the coefficient $c_2$ and the constant $a$ in \eqref{eq:q-holo-dict}.

It will be very interesting to establish the validity of \eqref{eq:logZ} for other three-manifolds using supersymmetric localization.  Another worthy goal is to extend the results for the partition function in \eqref{eq:logZ} and Table~\ref{tab:coeff} to other three-dimensional $\mathcal{N}=2$ SCFTs with a holographic dual in string or M-theory.  Prime candidates for such a generalization are Chern-Simons matter theories arising from M2-branes probing Sasaki-Einstein singularities. In addition, it should be possible to apply our results also to theories arising from D2-branes \cite{Guarino:2015jca} and wrapped D4-branes \cite{Crichigno:2018adf,Hosseini:2018uzp} in massive IIA string theory as well as M5-branes wrapped on three-manifolds in M-theory \cite{Gang:2018hjd}. More generally, the universality of the on-shell action result in \eqref{eq:i4} and its implications for the partition function of the dual SCFT underscores the ``unreasonable effectiveness" of our results and can be viewed as a four-derivative extension of the ideas discussed in \cite{Bobev:2017uzs}. 
%
%
\section*{Acknowledgements}
We are grateful to S. Chester, M. Crichigno, F.F. Gautason, S. Katmadas, B. McPeak, V. Min, and S. Pufu for useful discussions. NB is supported in part by an Odysseus grant G0F9516N from the FWO. AMC is supported in part by the European Research Council grant no. ERC-2013-CoG 616732 HoloQosmos. KH is supported in part by the Bulgarian NSF grants DN08/3, N28/5, and KP-06-N 38/11. NB, AMC, and VR are also supported by the KU Leuven C1 grant ZKD1118 C16/16/005.

\bibliography{HD-short}

\begin{thebibliography}{47}%
\makeatletter
\providecommand \@ifxundefined [1]{%
 \@ifx{#1\undefined}
}%
\providecommand \@ifnum [1]{%
 \ifnum #1\expandafter \@firstoftwo
 \else \expandafter \@secondoftwo
 \fi
}%
\providecommand \@ifx [1]{%
 \ifx #1\expandafter \@firstoftwo
 \else \expandafter \@secondoftwo
 \fi
}%
\providecommand \natexlab [1]{#1}%
\providecommand \enquote  [1]{``#1''}%
\providecommand \bibnamefont  [1]{#1}%
\providecommand \bibfnamefont [1]{#1}%
\providecommand \citenamefont [1]{#1}%
\providecommand \href@noop [0]{\@secondoftwo}%
\providecommand \href [0]{\begingroup \@sanitize@url \@href}%
\providecommand \@href[1]{\@@startlink{#1}\@@href}%
\providecommand \@@href[1]{\endgroup#1\@@endlink}%
\providecommand \@sanitize@url [0]{\catcode `\\12\catcode `\$12\catcode
  `\&12\catcode `\#12\catcode `\^12\catcode `\_12\catcode `\%12\relax}%
\providecommand \@@startlink[1]{}%
\providecommand \@@endlink[0]{}%
\providecommand \url  [0]{\begingroup\@sanitize@url \@url }%
\providecommand \@url [1]{\endgroup\@href {#1}{\urlprefix }}%
\providecommand \urlprefix  [0]{URL }%
\providecommand \Eprint [0]{\href }%
\providecommand \doibase [0]{http://dx.doi.org/}%
\providecommand \selectlanguage [0]{\@gobble}%
\providecommand \bibinfo  [0]{\@secondoftwo}%
\providecommand \bibfield  [0]{\@secondoftwo}%
\providecommand \translation [1]{[#1]}%
\providecommand \BibitemOpen [0]{}%
\providecommand \bibitemStop [0]{}%
\providecommand \bibitemNoStop [0]{.\EOS\space}%
\providecommand \EOS [0]{\spacefactor3000\relax}%
\providecommand \BibitemShut  [1]{\csname bibitem#1\endcsname}%
\let\auto@bib@innerbib\@empty
\bibitem [{\citenamefont {Lauria}\ and\ \citenamefont
  {Van~Proeyen}(2020)}]{Lauria:2020rhc}%
  \BibitemOpen
  \bibfield  {author} {\bibinfo {author} {\bibfnamefont {E.}~\bibnamefont
  {Lauria}}\ and\ \bibinfo {author} {\bibfnamefont {A.}~\bibnamefont
  {Van~Proeyen}},\ }\href {\doibase 10.1007/978-3-030-33757-5} {\emph {\bibinfo
  {title} {{${\cal N}=2$ Supergravity in $D=4,5,6$ Dimensions}}}},\ Vol.\
  \bibinfo {volume} {966}\ (\bibinfo  {publisher} {Springer},\ \bibinfo {year}
  {2020})\ \Eprint {http://arxiv.org/abs/2004.11433} {arXiv:2004.11433
  [hep-th]} \BibitemShut {NoStop}%
\bibitem [{\citenamefont {de~Wit}\ and\ \citenamefont
  {Reys}(2017)}]{deWit:2017cle}%
  \BibitemOpen
  \bibfield  {author} {\bibinfo {author} {\bibfnamefont {B.}~\bibnamefont
  {de~Wit}}\ and\ \bibinfo {author} {\bibfnamefont {V.}~\bibnamefont {Reys}},\
  }\href {\doibase 10.1007/JHEP12(2017)011} {\bibfield  {journal} {\bibinfo
  {journal} {JHEP}\ }\textbf {\bibinfo {volume} {12}},\ \bibinfo {pages} {011}
  (\bibinfo {year} {2017})},\ \Eprint {http://arxiv.org/abs/1706.04973}
  {arXiv:1706.04973 [hep-th]} \BibitemShut {NoStop}%
\bibitem [{\citenamefont {Butter}(2011)}]{Butter:2011sr}%
  \BibitemOpen
  \bibfield  {author} {\bibinfo {author} {\bibfnamefont {D.}~\bibnamefont
  {Butter}},\ }\href {\doibase 10.1007/JHEP10(2011)030} {\bibfield  {journal}
  {\bibinfo  {journal} {JHEP}\ }\textbf {\bibinfo {volume} {10}},\ \bibinfo
  {pages} {030} (\bibinfo {year} {2011})},\ \Eprint
  {http://arxiv.org/abs/1103.5914} {arXiv:1103.5914 [hep-th]} \BibitemShut
  {NoStop}%
\bibitem [{\citenamefont {Bergshoeff}\ \emph {et~al.}(1981)\citenamefont
  {Bergshoeff}, \citenamefont {de~Roo},\ and\ \citenamefont
  {de~Wit}}]{Bergshoeff:1980is}%
  \BibitemOpen
  \bibfield  {author} {\bibinfo {author} {\bibfnamefont {E.}~\bibnamefont
  {Bergshoeff}}, \bibinfo {author} {\bibfnamefont {M.}~\bibnamefont {de~Roo}},
  \ and\ \bibinfo {author} {\bibfnamefont {B.}~\bibnamefont {de~Wit}},\ }\href
  {\doibase 10.1016/0550-3213(81)90465-X} {\bibfield  {journal} {\bibinfo
  {journal} {Nucl. Phys. B}\ }\textbf {\bibinfo {volume} {182}},\ \bibinfo
  {pages} {173} (\bibinfo {year} {1981})}\BibitemShut {NoStop}%
\bibitem [{\citenamefont {Butter}\ \emph {et~al.}(2013)\citenamefont {Butter},
  \citenamefont {de~Wit}, \citenamefont {Kuzenko},\ and\ \citenamefont
  {Lodato}}]{Butter:2013lta}%
  \BibitemOpen
  \bibfield  {author} {\bibinfo {author} {\bibfnamefont {D.}~\bibnamefont
  {Butter}}, \bibinfo {author} {\bibfnamefont {B.}~\bibnamefont {de~Wit}},
  \bibinfo {author} {\bibfnamefont {S.~M.}\ \bibnamefont {Kuzenko}}, \ and\
  \bibinfo {author} {\bibfnamefont {I.}~\bibnamefont {Lodato}},\ }\href
  {\doibase 10.1007/JHEP12(2013)062} {\bibfield  {journal} {\bibinfo  {journal}
  {JHEP}\ }\textbf {\bibinfo {volume} {12}},\ \bibinfo {pages} {062} (\bibinfo
  {year} {2013})},\ \Eprint {http://arxiv.org/abs/1307.6546} {arXiv:1307.6546
  [hep-th]} \BibitemShut {NoStop}%
\bibitem [{\citenamefont {Bobev}\ \emph {et~al.}()\citenamefont {Bobev},
  \citenamefont {Charles}, \citenamefont {Hristov},\ and\ \citenamefont
  {Reys}}]{bigpaper}%
  \BibitemOpen
  \bibfield  {author} {\bibinfo {author} {\bibfnamefont {N.}~\bibnamefont
  {Bobev}}, \bibinfo {author} {\bibfnamefont {A.~M.}\ \bibnamefont {Charles}},
  \bibinfo {author} {\bibfnamefont {K.}~\bibnamefont {Hristov}}, \ and\
  \bibinfo {author} {\bibfnamefont {V.}~\bibnamefont {Reys}},\ }\href@noop {}
  {\bibinfo  {journal} {to appear}\ }\BibitemShut {NoStop}%
\bibitem [{\citenamefont {Charles}\ and\ \citenamefont
  {Larsen}(2016)}]{Charles:2016wjs}%
  \BibitemOpen
\bibfield  {journal} {  }\bibfield  {author} {\bibinfo {author} {\bibfnamefont
  {A.~M.}\ \bibnamefont {Charles}}\ and\ \bibinfo {author} {\bibfnamefont
  {F.}~\bibnamefont {Larsen}},\ }\href {\doibase 10.1007/JHEP10(2016)142}
  {\bibfield  {journal} {\bibinfo  {journal} {JHEP}\ }\textbf {\bibinfo
  {volume} {10}},\ \bibinfo {pages} {142} (\bibinfo {year} {2016})},\ \Eprint
  {http://arxiv.org/abs/1605.07622} {arXiv:1605.07622 [hep-th]} \BibitemShut
  {NoStop}%
\bibitem [{\citenamefont {Charles}\ \emph {et~al.}(2017)\citenamefont
  {Charles}, \citenamefont {Larsen},\ and\ \citenamefont
  {Mayerson}}]{Charles:2017dbr}%
  \BibitemOpen
  \bibfield  {author} {\bibinfo {author} {\bibfnamefont {A.~M.}\ \bibnamefont
  {Charles}}, \bibinfo {author} {\bibfnamefont {F.}~\bibnamefont {Larsen}}, \
  and\ \bibinfo {author} {\bibfnamefont {D.~R.}\ \bibnamefont {Mayerson}},\
  }\href {\doibase 10.1007/JHEP08(2017)048} {\bibfield  {journal} {\bibinfo
  {journal} {JHEP}\ }\textbf {\bibinfo {volume} {08}},\ \bibinfo {pages} {048}
  (\bibinfo {year} {2017})},\ \Eprint {http://arxiv.org/abs/1702.08458}
  {arXiv:1702.08458 [hep-th]} \BibitemShut {NoStop}%
\bibitem [{\citenamefont {Smolic}\ and\ \citenamefont
  {Taylor}(2013)}]{Smolic:2013gz}%
  \BibitemOpen
  \bibfield  {author} {\bibinfo {author} {\bibfnamefont {J.}~\bibnamefont
  {Smolic}}\ and\ \bibinfo {author} {\bibfnamefont {M.}~\bibnamefont
  {Taylor}},\ }\href {\doibase 10.1007/JHEP06(2013)096} {\bibfield  {journal}
  {\bibinfo  {journal} {JHEP}\ }\textbf {\bibinfo {volume} {06}},\ \bibinfo
  {pages} {096} (\bibinfo {year} {2013})},\ \Eprint
  {http://arxiv.org/abs/1301.5205} {arXiv:1301.5205 [hep-th]} \BibitemShut
  {NoStop}%
\bibitem [{\citenamefont {Emparan}\ \emph {et~al.}(1999)\citenamefont
  {Emparan}, \citenamefont {Johnson},\ and\ \citenamefont
  {Myers}}]{Emparan:1999pm}%
  \BibitemOpen
  \bibfield  {author} {\bibinfo {author} {\bibfnamefont {R.}~\bibnamefont
  {Emparan}}, \bibinfo {author} {\bibfnamefont {C.~V.}\ \bibnamefont
  {Johnson}}, \ and\ \bibinfo {author} {\bibfnamefont {R.~C.}\ \bibnamefont
  {Myers}},\ }\href {\doibase 10.1103/PhysRevD.60.104001} {\bibfield  {journal}
  {\bibinfo  {journal} {Phys. Rev. D}\ }\textbf {\bibinfo {volume} {60}},\
  \bibinfo {pages} {104001} (\bibinfo {year} {1999})},\ \Eprint
  {http://arxiv.org/abs/hep-th/9903238} {arXiv:hep-th/9903238} \BibitemShut
  {NoStop}%
\bibitem [{\citenamefont {Myers}(1987)}]{Myers:1987yn}%
  \BibitemOpen
  \bibfield  {author} {\bibinfo {author} {\bibfnamefont {R.~C.}\ \bibnamefont
  {Myers}},\ }\href {\doibase 10.1103/PhysRevD.36.392} {\bibfield  {journal}
  {\bibinfo  {journal} {Phys. Rev. D}\ }\textbf {\bibinfo {volume} {36}},\
  \bibinfo {pages} {392} (\bibinfo {year} {1987})}\BibitemShut {NoStop}%
\bibitem [{\citenamefont {Chamblin}\ \emph {et~al.}(1999)\citenamefont
  {Chamblin}, \citenamefont {Emparan}, \citenamefont {Johnson},\ and\
  \citenamefont {Myers}}]{Chamblin:1998pz}%
  \BibitemOpen
  \bibfield  {author} {\bibinfo {author} {\bibfnamefont {A.}~\bibnamefont
  {Chamblin}}, \bibinfo {author} {\bibfnamefont {R.}~\bibnamefont {Emparan}},
  \bibinfo {author} {\bibfnamefont {C.~V.}\ \bibnamefont {Johnson}}, \ and\
  \bibinfo {author} {\bibfnamefont {R.~C.}\ \bibnamefont {Myers}},\ }\href
  {\doibase 10.1103/PhysRevD.59.064010} {\bibfield  {journal} {\bibinfo
  {journal} {Phys. Rev. D}\ }\textbf {\bibinfo {volume} {59}},\ \bibinfo
  {pages} {064010} (\bibinfo {year} {1999})},\ \Eprint
  {http://arxiv.org/abs/hep-th/9808177} {arXiv:hep-th/9808177} \BibitemShut
  {NoStop}%
\bibitem [{\citenamefont {Martelli}\ \emph {et~al.}(2012)\citenamefont
  {Martelli}, \citenamefont {Passias},\ and\ \citenamefont
  {Sparks}}]{Martelli:2011fu}%
  \BibitemOpen
  \bibfield  {author} {\bibinfo {author} {\bibfnamefont {D.}~\bibnamefont
  {Martelli}}, \bibinfo {author} {\bibfnamefont {A.}~\bibnamefont {Passias}}, \
  and\ \bibinfo {author} {\bibfnamefont {J.}~\bibnamefont {Sparks}},\ }\href
  {\doibase 10.1016/j.nuclphysb.2012.07.019} {\bibfield  {journal} {\bibinfo
  {journal} {Nucl. Phys. B}\ }\textbf {\bibinfo {volume} {864}},\ \bibinfo
  {pages} {840} (\bibinfo {year} {2012})},\ \Eprint
  {http://arxiv.org/abs/1110.6400} {arXiv:1110.6400 [hep-th]} \BibitemShut
  {NoStop}%
\bibitem [{\citenamefont {Martelli}\ and\ \citenamefont
  {Sparks}(2013)}]{Martelli:2011fw}%
  \BibitemOpen
  \bibfield  {author} {\bibinfo {author} {\bibfnamefont {D.}~\bibnamefont
  {Martelli}}\ and\ \bibinfo {author} {\bibfnamefont {J.}~\bibnamefont
  {Sparks}},\ }\href {\doibase 10.1016/j.nuclphysb.2012.08.015} {\bibfield
  {journal} {\bibinfo  {journal} {Nucl. Phys. B}\ }\textbf {\bibinfo {volume}
  {866}},\ \bibinfo {pages} {72} (\bibinfo {year} {2013})},\ \Eprint
  {http://arxiv.org/abs/1111.6930} {arXiv:1111.6930 [hep-th]} \BibitemShut
  {NoStop}%
\bibitem [{\citenamefont {Farquet}\ \emph {et~al.}(2016)\citenamefont
  {Farquet}, \citenamefont {Lorenzen}, \citenamefont {Martelli},\ and\
  \citenamefont {Sparks}}]{Farquet:2014kma}%
  \BibitemOpen
  \bibfield  {author} {\bibinfo {author} {\bibfnamefont {D.}~\bibnamefont
  {Farquet}}, \bibinfo {author} {\bibfnamefont {J.}~\bibnamefont {Lorenzen}},
  \bibinfo {author} {\bibfnamefont {D.}~\bibnamefont {Martelli}}, \ and\
  \bibinfo {author} {\bibfnamefont {J.}~\bibnamefont {Sparks}},\ }\href
  {\doibase 10.1007/JHEP08(2016)080} {\bibfield  {journal} {\bibinfo  {journal}
  {JHEP}\ }\textbf {\bibinfo {volume} {08}},\ \bibinfo {pages} {080} (\bibinfo
  {year} {2016})},\ \Eprint {http://arxiv.org/abs/1404.0268} {arXiv:1404.0268
  [hep-th]} \BibitemShut {NoStop}%
\bibitem [{\citenamefont {Caldarelli}\ and\ \citenamefont
  {Klemm}(1999)}]{Caldarelli:1998hg}%
  \BibitemOpen
  \bibfield  {author} {\bibinfo {author} {\bibfnamefont {M.~M.}\ \bibnamefont
  {Caldarelli}}\ and\ \bibinfo {author} {\bibfnamefont {D.}~\bibnamefont
  {Klemm}},\ }\href {\doibase 10.1016/S0550-3213(98)00846-3} {\bibfield
  {journal} {\bibinfo  {journal} {Nucl. Phys. B}\ }\textbf {\bibinfo {volume}
  {545}},\ \bibinfo {pages} {434} (\bibinfo {year} {1999})},\ \Eprint
  {http://arxiv.org/abs/hep-th/9808097} {arXiv:hep-th/9808097} \BibitemShut
  {NoStop}%
\bibitem [{\citenamefont {Bobev}\ \emph {et~al.}(2020)\citenamefont {Bobev},
  \citenamefont {Charles},\ and\ \citenamefont {Min}}]{Bobev:2020pjk}%
  \BibitemOpen
  \bibfield  {author} {\bibinfo {author} {\bibfnamefont {N.}~\bibnamefont
  {Bobev}}, \bibinfo {author} {\bibfnamefont {A.~M.}\ \bibnamefont {Charles}},
  \ and\ \bibinfo {author} {\bibfnamefont {V.~S.}\ \bibnamefont {Min}},\
  }\href@noop {} {\  (\bibinfo {year} {2020})},\ \Eprint
  {http://arxiv.org/abs/2006.01148} {arXiv:2006.01148 [hep-th]} \BibitemShut
  {NoStop}%
\bibitem [{\citenamefont {Toldo}\ and\ \citenamefont
  {Willett}(2018)}]{Toldo:2017qsh}%
  \BibitemOpen
  \bibfield  {author} {\bibinfo {author} {\bibfnamefont {C.}~\bibnamefont
  {Toldo}}\ and\ \bibinfo {author} {\bibfnamefont {B.}~\bibnamefont
  {Willett}},\ }\href {\doibase 10.1007/JHEP05(2018)116} {\bibfield  {journal}
  {\bibinfo  {journal} {JHEP}\ }\textbf {\bibinfo {volume} {05}},\ \bibinfo
  {pages} {116} (\bibinfo {year} {2018})},\ \Eprint
  {http://arxiv.org/abs/1712.08861} {arXiv:1712.08861 [hep-th]} \BibitemShut
  {NoStop}%
\bibitem [{\citenamefont {Benetti~Genolini}\ \emph {et~al.}(2019)\citenamefont
  {Benetti~Genolini}, \citenamefont {Perez Ipi\~na},\ and\ \citenamefont
  {Sparks}}]{BenettiGenolini:2019jdz}%
  \BibitemOpen
  \bibfield  {author} {\bibinfo {author} {\bibfnamefont {P.}~\bibnamefont
  {Benetti~Genolini}}, \bibinfo {author} {\bibfnamefont {J.~M.}\ \bibnamefont
  {Perez Ipi\~na}}, \ and\ \bibinfo {author} {\bibfnamefont {J.}~\bibnamefont
  {Sparks}},\ }\href {\doibase 10.1007/JHEP10(2019)252} {\bibfield  {journal}
  {\bibinfo  {journal} {JHEP}\ }\textbf {\bibinfo {volume} {10}},\ \bibinfo
  {pages} {252} (\bibinfo {year} {2019})},\ \Eprint
  {http://arxiv.org/abs/1906.11249} {arXiv:1906.11249 [hep-th]} \BibitemShut
  {NoStop}%
\bibitem [{\citenamefont {Sen}\ and\ \citenamefont
  {Sinha}(2014)}]{Sen:2014nfa}%
  \BibitemOpen
  \bibfield  {author} {\bibinfo {author} {\bibfnamefont {K.}~\bibnamefont
  {Sen}}\ and\ \bibinfo {author} {\bibfnamefont {A.}~\bibnamefont {Sinha}},\
  }\href {\doibase 10.1007/JHEP07(2014)098} {\bibfield  {journal} {\bibinfo
  {journal} {JHEP}\ }\textbf {\bibinfo {volume} {07}},\ \bibinfo {pages} {098}
  (\bibinfo {year} {2014})},\ \Eprint {http://arxiv.org/abs/1405.7862}
  {arXiv:1405.7862 [hep-th]} \BibitemShut {NoStop}%
\bibitem [{\citenamefont {Chester}\ \emph {et~al.}(2020)\citenamefont
  {Chester}, \citenamefont {Kalloor},\ and\ \citenamefont
  {Sharon}}]{Chester:2020jay}%
  \BibitemOpen
  \bibfield  {author} {\bibinfo {author} {\bibfnamefont {S.~M.}\ \bibnamefont
  {Chester}}, \bibinfo {author} {\bibfnamefont {R.~R.}\ \bibnamefont
  {Kalloor}}, \ and\ \bibinfo {author} {\bibfnamefont {A.}~\bibnamefont
  {Sharon}},\ }\href@noop {} {\  (\bibinfo {year} {2020})},\ \Eprint
  {http://arxiv.org/abs/2004.13603} {arXiv:2004.13603 [hep-th]} \BibitemShut
  {NoStop}%
\bibitem [{\citenamefont {Closset}\ \emph {et~al.}(2013)\citenamefont
  {Closset}, \citenamefont {Dumitrescu}, \citenamefont {Festuccia},\ and\
  \citenamefont {Komargodski}}]{Closset:2012ru}%
  \BibitemOpen
  \bibfield  {author} {\bibinfo {author} {\bibfnamefont {C.}~\bibnamefont
  {Closset}}, \bibinfo {author} {\bibfnamefont {T.~T.}\ \bibnamefont
  {Dumitrescu}}, \bibinfo {author} {\bibfnamefont {G.}~\bibnamefont
  {Festuccia}}, \ and\ \bibinfo {author} {\bibfnamefont {Z.}~\bibnamefont
  {Komargodski}},\ }\href {\doibase 10.1007/JHEP05(2013)017} {\bibfield
  {journal} {\bibinfo  {journal} {JHEP}\ }\textbf {\bibinfo {volume} {05}},\
  \bibinfo {pages} {017} (\bibinfo {year} {2013})},\ \Eprint
  {http://arxiv.org/abs/1212.3388} {arXiv:1212.3388 [hep-th]} \BibitemShut
  {NoStop}%
\bibitem [{\citenamefont {Wald}(1993)}]{Wald:1993nt}%
  \BibitemOpen
  \bibfield  {author} {\bibinfo {author} {\bibfnamefont {R.~M.}\ \bibnamefont
  {Wald}},\ }\href {\doibase 10.1103/PhysRevD.48.R3427} {\bibfield  {journal}
  {\bibinfo  {journal} {Phys. Rev. D}\ }\textbf {\bibinfo {volume} {48}},\
  \bibinfo {pages} {3427} (\bibinfo {year} {1993})},\ \Eprint
  {http://arxiv.org/abs/gr-qc/9307038} {arXiv:gr-qc/9307038} \BibitemShut
  {NoStop}%
\bibitem [{\citenamefont {Susskind}\ and\ \citenamefont
  {Uglum}(1994)}]{Susskind:1994sm}%
  \BibitemOpen
  \bibfield  {author} {\bibinfo {author} {\bibfnamefont {L.}~\bibnamefont
  {Susskind}}\ and\ \bibinfo {author} {\bibfnamefont {J.}~\bibnamefont
  {Uglum}},\ }\href {\doibase 10.1103/PhysRevD.50.2700} {\bibfield  {journal}
  {\bibinfo  {journal} {Phys. Rev. D}\ }\textbf {\bibinfo {volume} {50}},\
  \bibinfo {pages} {2700} (\bibinfo {year} {1994})},\ \Eprint
  {http://arxiv.org/abs/hep-th/9401070} {arXiv:hep-th/9401070} \BibitemShut
  {NoStop}%
\bibitem [{\citenamefont {Larsen}\ and\ \citenamefont
  {Wilczek}(1996)}]{Larsen:1995ax}%
  \BibitemOpen
  \bibfield  {author} {\bibinfo {author} {\bibfnamefont {F.}~\bibnamefont
  {Larsen}}\ and\ \bibinfo {author} {\bibfnamefont {F.}~\bibnamefont
  {Wilczek}},\ }\href {\doibase 10.1016/0550-3213(95)00548-X} {\bibfield
  {journal} {\bibinfo  {journal} {Nucl. Phys. B}\ }\textbf {\bibinfo {volume}
  {458}},\ \bibinfo {pages} {249} (\bibinfo {year} {1996})},\ \Eprint
  {http://arxiv.org/abs/hep-th/9506066} {arXiv:hep-th/9506066} \BibitemShut
  {NoStop}%
\bibitem [{\citenamefont {Balasubramanian}\ and\ \citenamefont
  {Kraus}(1999)}]{Balasubramanian:1999re}%
  \BibitemOpen
  \bibfield  {author} {\bibinfo {author} {\bibfnamefont {V.}~\bibnamefont
  {Balasubramanian}}\ and\ \bibinfo {author} {\bibfnamefont {P.}~\bibnamefont
  {Kraus}},\ }\href {\doibase 10.1007/s002200050764} {\bibfield  {journal}
  {\bibinfo  {journal} {Commun. Math. Phys.}\ }\textbf {\bibinfo {volume}
  {208}},\ \bibinfo {pages} {413} (\bibinfo {year} {1999})},\ \Eprint
  {http://arxiv.org/abs/hep-th/9902121} {arXiv:hep-th/9902121} \BibitemShut
  {NoStop}%
\bibitem [{\citenamefont {Myers}\ and\ \citenamefont
  {Simon}(1988)}]{Myers:1988ze}%
  \BibitemOpen
  \bibfield  {author} {\bibinfo {author} {\bibfnamefont {R.~C.}\ \bibnamefont
  {Myers}}\ and\ \bibinfo {author} {\bibfnamefont {J.~Z.}\ \bibnamefont
  {Simon}},\ }\href {\doibase 10.1103/PhysRevD.38.2434} {\bibfield  {journal}
  {\bibinfo  {journal} {Phys.\ Rev.\ D}\ }\textbf {\bibinfo {volume} {38}},\
  \bibinfo {pages} {2434} (\bibinfo {year} {1988})}\BibitemShut {NoStop}%
\bibitem [{\citenamefont {Gibbons}\ \emph {et~al.}(2005)\citenamefont
  {Gibbons}, \citenamefont {Perry},\ and\ \citenamefont
  {Pope}}]{Gibbons:2004ai}%
  \BibitemOpen
  \bibfield  {author} {\bibinfo {author} {\bibfnamefont {G.}~\bibnamefont
  {Gibbons}}, \bibinfo {author} {\bibfnamefont {M.}~\bibnamefont {Perry}}, \
  and\ \bibinfo {author} {\bibfnamefont {C.}~\bibnamefont {Pope}},\ }\href
  {\doibase 10.1088/0264-9381/22/9/002} {\bibfield  {journal} {\bibinfo
  {journal} {Class. Quant. Grav.}\ }\textbf {\bibinfo {volume} {22}},\ \bibinfo
  {pages} {1503} (\bibinfo {year} {2005})},\ \Eprint
  {http://arxiv.org/abs/hep-th/0408217} {arXiv:hep-th/0408217} \BibitemShut
  {NoStop}%
\bibitem [{\citenamefont {Cheung}\ \emph {et~al.}(2018)\citenamefont {Cheung},
  \citenamefont {Liu},\ and\ \citenamefont {Remmen}}]{Cheung:2018cwt}%
  \BibitemOpen
  \bibfield  {author} {\bibinfo {author} {\bibfnamefont {C.}~\bibnamefont
  {Cheung}}, \bibinfo {author} {\bibfnamefont {J.}~\bibnamefont {Liu}}, \ and\
  \bibinfo {author} {\bibfnamefont {G.~N.}\ \bibnamefont {Remmen}},\ }\href
  {\doibase 10.1007/JHEP10(2018)004} {\bibfield  {journal} {\bibinfo  {journal}
  {JHEP}\ }\textbf {\bibinfo {volume} {10}},\ \bibinfo {pages} {004} (\bibinfo
  {year} {2018})},\ \Eprint {http://arxiv.org/abs/1801.08546} {arXiv:1801.08546
  [hep-th]} \BibitemShut {NoStop}%
\bibitem [{\citenamefont {Goon}\ and\ \citenamefont
  {Penco}(2020)}]{Goon:2019faz}%
  \BibitemOpen
  \bibfield  {author} {\bibinfo {author} {\bibfnamefont {G.}~\bibnamefont
  {Goon}}\ and\ \bibinfo {author} {\bibfnamefont {R.}~\bibnamefont {Penco}},\
  }\href {\doibase 10.1103/PhysRevLett.124.101103} {\bibfield  {journal}
  {\bibinfo  {journal} {Phys. Rev. Lett.}\ }\textbf {\bibinfo {volume} {124}},\
  \bibinfo {pages} {101103} (\bibinfo {year} {2020})},\ \Eprint
  {http://arxiv.org/abs/1909.05254} {arXiv:1909.05254 [hep-th]} \BibitemShut
  {NoStop}%
\bibitem [{\citenamefont {Cremonini}\ \emph {et~al.}(2019)\citenamefont
  {Cremonini}, \citenamefont {Jones}, \citenamefont {Liu},\ and\ \citenamefont
  {McPeak}}]{Cremonini:2019wdk}%
  \BibitemOpen
  \bibfield  {author} {\bibinfo {author} {\bibfnamefont {S.}~\bibnamefont
  {Cremonini}}, \bibinfo {author} {\bibfnamefont {C.~R.}\ \bibnamefont
  {Jones}}, \bibinfo {author} {\bibfnamefont {J.~T.}\ \bibnamefont {Liu}}, \
  and\ \bibinfo {author} {\bibfnamefont {B.}~\bibnamefont {McPeak}},\
  }\href@noop {} {\  (\bibinfo {year} {2019})},\ \Eprint
  {http://arxiv.org/abs/1912.11161} {arXiv:1912.11161 [hep-th]} \BibitemShut
  {NoStop}%
\bibitem [{\citenamefont {Gauntlett}\ and\ \citenamefont
  {Varela}(2007)}]{Gauntlett:2007ma}%
  \BibitemOpen
  \bibfield  {author} {\bibinfo {author} {\bibfnamefont {J.~P.}\ \bibnamefont
  {Gauntlett}}\ and\ \bibinfo {author} {\bibfnamefont {O.}~\bibnamefont
  {Varela}},\ }\href {\doibase 10.1103/PhysRevD.76.126007} {\bibfield
  {journal} {\bibinfo  {journal} {Phys. Rev. D}\ }\textbf {\bibinfo {volume}
  {76}},\ \bibinfo {pages} {126007} (\bibinfo {year} {2007})},\ \Eprint
  {http://arxiv.org/abs/0707.2315} {arXiv:0707.2315 [hep-th]} \BibitemShut
  {NoStop}%
\bibitem [{\citenamefont {Aharony}\ \emph {et~al.}(2008)\citenamefont
  {Aharony}, \citenamefont {Bergman}, \citenamefont {Jafferis},\ and\
  \citenamefont {Maldacena}}]{Aharony:2008ug}%
  \BibitemOpen
  \bibfield  {author} {\bibinfo {author} {\bibfnamefont {O.}~\bibnamefont
  {Aharony}}, \bibinfo {author} {\bibfnamefont {O.}~\bibnamefont {Bergman}},
  \bibinfo {author} {\bibfnamefont {D.~L.}\ \bibnamefont {Jafferis}}, \ and\
  \bibinfo {author} {\bibfnamefont {J.}~\bibnamefont {Maldacena}},\ }\href
  {\doibase 10.1088/1126-6708/2008/10/091} {\bibfield  {journal} {\bibinfo
  {journal} {JHEP}\ }\textbf {\bibinfo {volume} {10}},\ \bibinfo {pages} {091}
  (\bibinfo {year} {2008})},\ \Eprint {http://arxiv.org/abs/0806.1218}
  {arXiv:0806.1218 [hep-th]} \BibitemShut {NoStop}%
\bibitem [{\citenamefont {Mezei}\ and\ \citenamefont
  {Pufu}(2014)}]{Mezei:2013gqa}%
  \BibitemOpen
  \bibfield  {author} {\bibinfo {author} {\bibfnamefont {M.}~\bibnamefont
  {Mezei}}\ and\ \bibinfo {author} {\bibfnamefont {S.~S.}\ \bibnamefont
  {Pufu}},\ }\href {\doibase 10.1007/JHEP02(2014)037} {\bibfield  {journal}
  {\bibinfo  {journal} {JHEP}\ }\textbf {\bibinfo {volume} {02}},\ \bibinfo
  {pages} {037} (\bibinfo {year} {2014})},\ \Eprint
  {http://arxiv.org/abs/1312.0920} {arXiv:1312.0920 [hep-th]} \BibitemShut
  {NoStop}%
\bibitem [{\citenamefont {Camanho}\ \emph {et~al.}(2016)\citenamefont
  {Camanho}, \citenamefont {Edelstein}, \citenamefont {Maldacena},\ and\
  \citenamefont {Zhiboedov}}]{Camanho:2014apa}%
  \BibitemOpen
  \bibfield  {author} {\bibinfo {author} {\bibfnamefont {X.~O.}\ \bibnamefont
  {Camanho}}, \bibinfo {author} {\bibfnamefont {J.~D.}\ \bibnamefont
  {Edelstein}}, \bibinfo {author} {\bibfnamefont {J.}~\bibnamefont
  {Maldacena}}, \ and\ \bibinfo {author} {\bibfnamefont {A.}~\bibnamefont
  {Zhiboedov}},\ }\href {\doibase 10.1007/JHEP02(2016)020} {\bibfield
  {journal} {\bibinfo  {journal} {JHEP}\ }\textbf {\bibinfo {volume} {02}},\
  \bibinfo {pages} {020} (\bibinfo {year} {2016})},\ \Eprint
  {http://arxiv.org/abs/1407.5597} {arXiv:1407.5597 [hep-th]} \BibitemShut
  {NoStop}%
\bibitem [{\citenamefont {Marino}\ and\ \citenamefont
  {Putrov}(2012)}]{Marino:2011eh}%
  \BibitemOpen
  \bibfield  {author} {\bibinfo {author} {\bibfnamefont {M.}~\bibnamefont
  {Marino}}\ and\ \bibinfo {author} {\bibfnamefont {P.}~\bibnamefont
  {Putrov}},\ }\href {\doibase 10.1088/1742-5468/2012/03/P03001} {\bibfield
  {journal} {\bibinfo  {journal} {J. Stat. Mech.}\ }\textbf {\bibinfo {volume}
  {1203}},\ \bibinfo {pages} {P03001} (\bibinfo {year} {2012})},\ \Eprint
  {http://arxiv.org/abs/1110.4066} {arXiv:1110.4066 [hep-th]} \BibitemShut
  {NoStop}%
\bibitem [{\citenamefont {Fuji}\ \emph {et~al.}(2011)\citenamefont {Fuji},
  \citenamefont {Hirano},\ and\ \citenamefont {Moriyama}}]{Fuji:2011km}%
  \BibitemOpen
  \bibfield  {author} {\bibinfo {author} {\bibfnamefont {H.}~\bibnamefont
  {Fuji}}, \bibinfo {author} {\bibfnamefont {S.}~\bibnamefont {Hirano}}, \ and\
  \bibinfo {author} {\bibfnamefont {S.}~\bibnamefont {Moriyama}},\ }\href
  {\doibase 10.1007/JHEP08(2011)001} {\bibfield  {journal} {\bibinfo  {journal}
  {JHEP}\ }\textbf {\bibinfo {volume} {08}},\ \bibinfo {pages} {001} (\bibinfo
  {year} {2011})},\ \Eprint {http://arxiv.org/abs/1106.4631} {arXiv:1106.4631
  [hep-th]} \BibitemShut {NoStop}%
\bibitem [{\citenamefont {Hatsuda}(2016)}]{Hatsuda:2016uqa}%
  \BibitemOpen
  \bibfield  {author} {\bibinfo {author} {\bibfnamefont {Y.}~\bibnamefont
  {Hatsuda}},\ }\href {\doibase 10.1007/JHEP07(2016)026} {\bibfield  {journal}
  {\bibinfo  {journal} {JHEP}\ }\textbf {\bibinfo {volume} {07}},\ \bibinfo
  {pages} {026} (\bibinfo {year} {2016})},\ \Eprint
  {http://arxiv.org/abs/1601.02728} {arXiv:1601.02728 [hep-th]} \BibitemShut
  {NoStop}%
\bibitem [{\citenamefont {Benini}\ \emph {et~al.}(2016)\citenamefont {Benini},
  \citenamefont {Hristov},\ and\ \citenamefont {Zaffaroni}}]{Benini:2015eyy}%
  \BibitemOpen
  \bibfield  {author} {\bibinfo {author} {\bibfnamefont {F.}~\bibnamefont
  {Benini}}, \bibinfo {author} {\bibfnamefont {K.}~\bibnamefont {Hristov}}, \
  and\ \bibinfo {author} {\bibfnamefont {A.}~\bibnamefont {Zaffaroni}},\ }\href
  {\doibase 10.1007/JHEP05(2016)054} {\bibfield  {journal} {\bibinfo  {journal}
  {JHEP}\ }\textbf {\bibinfo {volume} {05}},\ \bibinfo {pages} {054} (\bibinfo
  {year} {2016})},\ \Eprint {http://arxiv.org/abs/1511.04085} {arXiv:1511.04085
  [hep-th]} \BibitemShut {NoStop}%
\bibitem [{\citenamefont {Azzurli}\ \emph {et~al.}(2018)\citenamefont
  {Azzurli}, \citenamefont {Bobev}, \citenamefont {Crichigno}, \citenamefont
  {Min},\ and\ \citenamefont {Zaffaroni}}]{Azzurli:2017kxo}%
  \BibitemOpen
  \bibfield  {author} {\bibinfo {author} {\bibfnamefont {F.}~\bibnamefont
  {Azzurli}}, \bibinfo {author} {\bibfnamefont {N.}~\bibnamefont {Bobev}},
  \bibinfo {author} {\bibfnamefont {P.~M.}\ \bibnamefont {Crichigno}}, \bibinfo
  {author} {\bibfnamefont {V.~S.}\ \bibnamefont {Min}}, \ and\ \bibinfo
  {author} {\bibfnamefont {A.}~\bibnamefont {Zaffaroni}},\ }\href {\doibase
  10.1007/JHEP02(2018)054} {\bibfield  {journal} {\bibinfo  {journal} {JHEP}\
  }\textbf {\bibinfo {volume} {02}},\ \bibinfo {pages} {054} (\bibinfo {year}
  {2018})},\ \Eprint {http://arxiv.org/abs/1707.04257} {arXiv:1707.04257
  [hep-th]} \BibitemShut {NoStop}%
\bibitem [{\citenamefont {Liu}\ \emph {et~al.}(2018)\citenamefont {Liu},
  \citenamefont {Pando~Zayas}, \citenamefont {Rathee},\ and\ \citenamefont
  {Zhao}}]{Liu:2017vll}%
  \BibitemOpen
  \bibfield  {author} {\bibinfo {author} {\bibfnamefont {J.~T.}\ \bibnamefont
  {Liu}}, \bibinfo {author} {\bibfnamefont {L.~A.}\ \bibnamefont
  {Pando~Zayas}}, \bibinfo {author} {\bibfnamefont {V.}~\bibnamefont {Rathee}},
  \ and\ \bibinfo {author} {\bibfnamefont {W.}~\bibnamefont {Zhao}},\ }\href
  {\doibase 10.1007/JHEP01(2018)026} {\bibfield  {journal} {\bibinfo  {journal}
  {JHEP}\ }\textbf {\bibinfo {volume} {01}},\ \bibinfo {pages} {026} (\bibinfo
  {year} {2018})},\ \Eprint {http://arxiv.org/abs/1707.04197} {arXiv:1707.04197
  [hep-th]} \BibitemShut {NoStop}%
\bibitem [{\citenamefont {Chester}\ \emph {et~al.}(2018)\citenamefont
  {Chester}, \citenamefont {Pufu},\ and\ \citenamefont
  {Yin}}]{Chester:2018aca}%
  \BibitemOpen
  \bibfield  {author} {\bibinfo {author} {\bibfnamefont {S.~M.}\ \bibnamefont
  {Chester}}, \bibinfo {author} {\bibfnamefont {S.~S.}\ \bibnamefont {Pufu}}, \
  and\ \bibinfo {author} {\bibfnamefont {X.}~\bibnamefont {Yin}},\ }\href
  {\doibase 10.1007/JHEP08(2018)115} {\bibfield  {journal} {\bibinfo  {journal}
  {JHEP}\ }\textbf {\bibinfo {volume} {08}},\ \bibinfo {pages} {115} (\bibinfo
  {year} {2018})},\ \Eprint {http://arxiv.org/abs/1804.00949} {arXiv:1804.00949
  [hep-th]} \BibitemShut {NoStop}%
\bibitem [{\citenamefont {Guarino}\ \emph {et~al.}(2015)\citenamefont
  {Guarino}, \citenamefont {Jafferis},\ and\ \citenamefont
  {Varela}}]{Guarino:2015jca}%
  \BibitemOpen
  \bibfield  {author} {\bibinfo {author} {\bibfnamefont {A.}~\bibnamefont
  {Guarino}}, \bibinfo {author} {\bibfnamefont {D.~L.}\ \bibnamefont
  {Jafferis}}, \ and\ \bibinfo {author} {\bibfnamefont {O.}~\bibnamefont
  {Varela}},\ }\href {\doibase 10.1103/PhysRevLett.115.091601} {\bibfield
  {journal} {\bibinfo  {journal} {Phys. Rev. Lett.}\ }\textbf {\bibinfo
  {volume} {115}},\ \bibinfo {pages} {091601} (\bibinfo {year} {2015})},\
  \Eprint {http://arxiv.org/abs/1504.08009} {arXiv:1504.08009 [hep-th]}
  \BibitemShut {NoStop}%
\bibitem [{\citenamefont {Crichigno}\ \emph {et~al.}(2018)\citenamefont
  {Crichigno}, \citenamefont {Jain},\ and\ \citenamefont
  {Willett}}]{Crichigno:2018adf}%
  \BibitemOpen
  \bibfield  {author} {\bibinfo {author} {\bibfnamefont {P.~M.}\ \bibnamefont
  {Crichigno}}, \bibinfo {author} {\bibfnamefont {D.}~\bibnamefont {Jain}}, \
  and\ \bibinfo {author} {\bibfnamefont {B.}~\bibnamefont {Willett}},\ }\href
  {\doibase 10.1007/JHEP11(2018)058} {\bibfield  {journal} {\bibinfo  {journal}
  {JHEP}\ }\textbf {\bibinfo {volume} {11}},\ \bibinfo {pages} {058} (\bibinfo
  {year} {2018})},\ \Eprint {http://arxiv.org/abs/1808.06744} {arXiv:1808.06744
  [hep-th]} \BibitemShut {NoStop}%
\bibitem [{\citenamefont {Hosseini}\ \emph {et~al.}(2018)\citenamefont
  {Hosseini}, \citenamefont {Yaakov},\ and\ \citenamefont
  {Zaffaroni}}]{Hosseini:2018uzp}%
  \BibitemOpen
  \bibfield  {author} {\bibinfo {author} {\bibfnamefont {S.~M.}\ \bibnamefont
  {Hosseini}}, \bibinfo {author} {\bibfnamefont {I.}~\bibnamefont {Yaakov}}, \
  and\ \bibinfo {author} {\bibfnamefont {A.}~\bibnamefont {Zaffaroni}},\ }\href
  {\doibase 10.1007/JHEP11(2018)119} {\bibfield  {journal} {\bibinfo  {journal}
  {JHEP}\ }\textbf {\bibinfo {volume} {11}},\ \bibinfo {pages} {119} (\bibinfo
  {year} {2018})},\ \Eprint {http://arxiv.org/abs/1808.06626} {arXiv:1808.06626
  [hep-th]} \BibitemShut {NoStop}%
\bibitem [{\citenamefont {Gang}\ and\ \citenamefont
  {Kim}(2019)}]{Gang:2018hjd}%
  \BibitemOpen
  \bibfield  {author} {\bibinfo {author} {\bibfnamefont {D.}~\bibnamefont
  {Gang}}\ and\ \bibinfo {author} {\bibfnamefont {N.}~\bibnamefont {Kim}},\
  }\href {\doibase 10.1103/PhysRevD.99.021901} {\bibfield  {journal} {\bibinfo
  {journal} {Phys. Rev. D}\ }\textbf {\bibinfo {volume} {99}},\ \bibinfo
  {pages} {021901} (\bibinfo {year} {2019})},\ \Eprint
  {http://arxiv.org/abs/1808.02797} {arXiv:1808.02797 [hep-th]} \BibitemShut
  {NoStop}%
\bibitem [{\citenamefont {Bobev}\ and\ \citenamefont
  {Crichigno}(2017)}]{Bobev:2017uzs}%
  \BibitemOpen
  \bibfield  {author} {\bibinfo {author} {\bibfnamefont {N.}~\bibnamefont
  {Bobev}}\ and\ \bibinfo {author} {\bibfnamefont {P.~M.}\ \bibnamefont
  {Crichigno}},\ }\href {\doibase 10.1007/JHEP12(2017)065} {\bibfield
  {journal} {\bibinfo  {journal} {JHEP}\ }\textbf {\bibinfo {volume} {12}},\
  \bibinfo {pages} {065} (\bibinfo {year} {2017})},\ \Eprint
  {http://arxiv.org/abs/1708.05052} {arXiv:1708.05052 [hep-th]} \BibitemShut
  {NoStop}%
\end{thebibliography}%

\end{document}